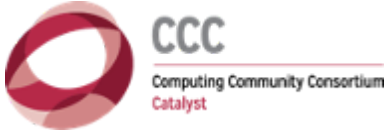

# Next Wave Artificial Intelligence: Robust, Explainable, Adaptable, Ethical, and Accountable

*A Computing Community Consortium (CCC) Quadrennial Paper*

*Odest Chadwicke Jenkins (University of Michigan), Daniel Lopresti (Lehigh University), and Melanie Mitchell (Portland State University and Santa Fe Institute)*

We are now seeing the impact of decades of investment in artificial intelligence (AI) across our society. In recent years, AI systems have been deployed in a broad array of application areas, including healthcare, transportation, finance, design and manufacturing, education, scientific discovery, and national security, among others. Many of these applications have addressed important societal problems and directly improved peoples' lives. However, broad beneficial use of AI applications are often stymied by the limitations of today's state-of-the-art systems. In this brief overview, we describe the limitations of today's AI systems and make recommendations for focus areas that will enable the field to move to the next level in terms of robustness and trustworthiness.

The history of AI has included several "waves" of ideas. The first wave, from the mid-1950s to the 1980s, focused on logic and symbolic hand-encoded representations of knowledge, the foundations of so-called "expert systems". The second wave, starting in the 1990s, focused on statistics and machine learning, in which, instead of hand-programming rules for behavior, programmers constructed "statistical learning algorithms" that could be trained on large datasets. In the most recent wave, especially in the last decade, research in AI has largely focused on deep (i.e., many-layered) neural networks, which are loosely inspired by the brain and trained by "deep learning" methods. However, while deep neural networks have led to many successes and new capabilities in computer vision, speech recognition, language processing, game-playing, and robotics, their potential for broad application remains limited by several factors. Deep neural networks typically require "supervised" training on large datasets—that is on thousands to millions of examples that have been manually labeled; these labeling efforts often require prohibitive amounts of human labor. Moreover, the labels can contain errors as well as both overt and subtle biases. Deep learning methods also require large computing infrastructure, whose electricity use can have negative environmental impacts.

A more concerning limitation is that even the most successful of today's AI systems suffer from *brittleness*—they can fail in unexpected ways when faced with situations that differ sufficiently from ones they have been trained on. This lack of robustness also appears in the vulnerability of AI systems to *adversarial attacks*, in which an adversary can subtly manipulate data in a way to guarantee a specific wrong answer or action from an AI system. AI systems also can absorb biases—based on gender, race, or other factors—from their training data and further magnify these biases in their subsequent decision-making. All of these issues are further complicated by the inherent complexity of today's state-of-the-art AI systems, which can often make their decision-making processes obscure, even to their own human designers. Taken together, these various limitations have prevented AI systems such as automatic medical diagnosis or autonomous vehicles from being sufficiently trustworthy for wide deployment. The massive proliferation of AI across society will require radically new ideas to yield technology that will not sacrifice our productivity, our quality of life, or our values. There are still major research challenges that we need to overcome to realize AI systems that will be robust, explainable, fast, mostly accurate, and adaptable to new circumstances.

In order to build robust, explainable, adaptable, and ethical AI systems, we recommend a substantial investment in new ideas—the "next wave"—that go beyond the current emphasis on supervised deep learning. The new ideas range from new thought about the fundamentals of AI, the relevance of discoveries from cognitive and brain science, as well as, revisiting the basic assumptions of computing, considering all options between digital and quantum. The United States Federal Government, as it has over the last century for computing[1], is in the unique position to invest in these new ideas, help them take shape, and benefit from the technological leadership likely to result.

### Recommendations for Catalyzing the Next Wave of AI

There is a new wave of AI ready to be pioneered that brings together the best of AI's many past advancements and extends this union towards the vision many have for AI. This new wave will coalesce the speed and recall power of today's data-driven AI with the robustness and explainability of traditional symbolic, logic-based AI, along with new ideas related to human commonsense reasoning, analogy-making, conceptual abstraction, and developmental learning.

### Integrating Existing Knowledge with Machine Learning

As discussed in the CCC AI Roadmap [AI19], there are significant benefits in finding ways to integrate knowledge sources (including existing scientific knowledge) with modern machine learning techniques. For example, many of the processes governing physics, chemistry, and biology are already well-characterized mathematically; these are sometimes called "first principles" and there is no need to "re-learn" them. Indeed, data-driven approaches that are oblivious to existing knowledge risk learning the wrong thing, especially when the available data is sparse relative to the dimensionality and the size of the problem space. On the other hand, for complex tasks critically important to society, such as modeling the world climate or the global economy, it may be impossible to develop complete mathematical models with closed-form solutions. This is where the power of New Wave AI may truly be felt. Here, the ability to seamlessly incorporate large quantities of real-world data to help calibrate the models and "fill in the gaps" will be vital. There is an appealing synergy, then, between existing knowledge and data driven AI; substantial research is needed to bring this to fruition.

The AI Roadmap includes a vignette describing a climate scientist seeking to make predictions at a variety of scales who needs to integrate mathematical models for atmospheric physics, solar radiation, and land surface-atmosphere interactions that, while extremely valuable, fall short without the inclusion of observational data; the two approaches complement one another, neither alone is sufficient [AI19]. (See also the companion CCC Quad Paper on the Rise of AI-Driven Simulators.)

---

[1] Waldrop, M. M. (2002). *The Dream Machine: J.C.R. Licklider and the Revolution That Made Computing Personal* (1st ed.). Penguin Books.

Another potential benefit of incorporating human knowledge with machine learning is reducing the resources needed to achieve a given level of performance. Deep neural networks are typically trained on massive datasets which require substantial human effort to gather and label. In addition, the algorithms for training DNNs are complex iterative procedures that consume significant processor cycles and hence large amounts of energy. The use of existing knowledge could reduce data and processing demands, making AI more sustainable. Moreover, because such knowledge arises from humans' attempts to understand natural phenomena, research in this area could also provide insights useful in developing explainable AI systems, another major goal of New Wave AI.

### Going Beyond Supervised Learning
Supervised learning, in which an algorithm (e.g., a deep neural network) learns from thousands to millions of human-labeled examples, has been the primary driver of most recent progress in AI. However, as we outlined at the beginning of this paper, supervised learning methods have many limitations. In order to make AI systems more robust, general, and trustworthy, the field needs to make progress on non-supervised learning methods. There is currently significant research activity on a variety of such methods, under rubrics such as "self-supervised learning", "reinforcement learning", and "active learning".

Some of the most promising research along these lines is inspired by the learning strategies of young children, who learn in an active and largely unsupervised manner in ways driven by both specific goals and open-ended curiosity. Perhaps most significantly, young children learn intricate, dynamic models of the world, ones that rely on knowledge of causality, rather than the kinds of statistical correlations that today's machine learning methods rely on. Such mental models allow for abstraction and analogy making—essential cognitive abilities that allow humans to transfer knowledge from one situation to related situations. Children's learning is bootstrapped by the innate or early-learned core knowledge humans share about the world. Such mental models form the heart of what we call "common sense", which is still missing from even the most advanced of today's AI systems. While machines can beat the best human chess and Go masters, paradoxically these same machines lack the common sense of a toddler. Research in how to give machines the core knowledge and reasoning abilities needed for common sense is needed to make possible the next wave of progress in AI.

### Integrating Probabilistic and Causal Models with Deep Learning
Probabilistic models such as Bayesian Networks have long played an important role in AI, especially in applications that require reasoning under uncertainty. However, such models are often difficult to learn from data, and require substantial human "knowledge engineering". We recommend expanded research on Integrating such models with deep learning systems; such research has the potential to yield hybrid models with abilities for both sophisticated pattern-recognition and rigorous probabilistic reasoning.

Furthermore, in order to imbue AI systems with humanlike commonsense reasoning—the ability to generalize and the ability to explain their own decisions— it will be essential to enable these systems to learn models that include *causality*, rather than their current reliance on statistical correlations. As stated in the CCC's AI Roadmap, "Advances in understanding causal inference could, for example, greatly advance research in biology and medicine, where these relationships are the key to designing medical interventions to fight disease."

**Incentivizing Ethical AI**
There is no denying that among the motivations driving AI research forward today, chief among them is the almost frenetic push in industry to be the first to develop and monetize a new product, while in academia a primary goal is the race to publish the "latest and greatest" results, whether theoretical or empirical. While there is now a growing awareness of the ethical implications of AI, the ramifications of AI solutions are often relegated to a second-tier consideration and cannot compete with profits or citations. Maintaining the status quo is clearly insufficient; as a result, there has been a flurry of activity in Washington, DC and elsewhere around the world to write new legislation to better govern the uses of AI, as well as the creation of industry partnerships and professional society working groups focused on the topic of ethical AI. There likewise needs to be funding for interdisciplinary research efforts in universities that bring together experts from a wide range of disciplines with complementary overlapping interests in AI ethics, including the social sciences, business, and law, in addition to computing. The Human-Centered AI Institute at Stanford, the AI Now Institute at NYU, and the Partnership on AI are examples of a growing array of such interdisciplinary efforts; such efforts should be expanded to many more institutions. Beyond a fundamental change to the mindset of those developing AI techniques, the rewards systems for fielding new products and publishing in the scientific literature must also evolve to incorporate ethics as a guiding principle, not an afterthought.

**Policies for Accountability and Liability**
When AI goes wrong, who should be held accountable? The theoretician who created the algorithm? The developer who wrote the code? The engineer who trained and tested the machine learning "black box"? The human factors expert who helped design the system? The company that integrated the AI software into its product? The human operator of the system who worked alongside the AI and was present, presumably, to catch any mistakes that arose and take control in time to avoid disaster?

The answer, right now at least, is "any or all of the above," although there seems to be an early tendency to attempt to assign blame to the human operator of the AI. While some of the important decisions will be made by lawmakers working alongside the insurance industry, which exists to quantify risk and make a profit by providing financial protection, computing researchers are uniquely well-positioned to understand the functioning and failings of complex software and software/hardware systems. This is another key area for interdisciplinary collaborations.

**Summary**

AI has achieved tremendous success at an accelerating pace over the past decade, but we are now entering a period where fundamental challenges will need to be addressed to unlock the next phase of advances. Here we have identified those that we consider most significant, including finding ways to integrate existing knowledge alongside data-hungry machine learning approaches, to move beyond the need to have human experts label large quantities of training data, to develop techniques that can learn causality and not only correlations, to create incentives to encourage ethical applications of AI, and to deal with open issues of accountability and liability as AI takes on more and more of the tasks previously performed by humans. Increased levels of funding to support research in these areas will be vital to moving AI forward.

**References**

[AI19] "A 20-Year Community Roadmap for Artificial Intelligence Research in the US," Yolanda Gil and Bart Selman, editors, Computing Community Consortium, 2019.

*This white paper is part of a series of papers compiled every four years by the CCC Council and members of the computing research community to inform policymakers, community members and the public on important research opportunities in areas of national priority. The topics chosen represent areas of pressing national need spanning various subdisciplines of the computing research field. The white papers attempt to portray a comprehensive picture of the computing research field detailing potential research directions, challenges and recommendations.*

*This material is based upon work supported by the National Science Foundation under Grant No. 1734706. Any opinions, findings, and conclusions or recommendations expressed in this material are those of the authors and do not necessarily reflect the views of the National Science Foundation.*

*For citation use: Jenkins C., Lopresti D. & Mitchell M. (2020) Next Wave Artificial Intelligence: Robust, Explainable, Adaptable, Ethical, and Accountable.*
*https://cra.org/ccc/resources/ccc-led-whitepapers/#2020-quadrennial-papers*